\documentclass[a4paper]{article}

\usepackage{graphicx,float}

\usepackage{multirow}
\usepackage{url}

\usepackage{comment}

\usepackage{todonotes}
\usepackage{amsthm,amsmath,amssymb,latexsym,stmaryrd,extarrows}
\usepackage[all]{xy}
\usepackage{color}
\usepackage{datetime}
\usepackage{fullpage}
\usepackage{mathrsfs}
\usepackage{xfrac}
\usepackage{bussproofs}
\usepackage{txfonts}

\usepackage{tikz}

\usepackage{times}

\usepackage{amsmath}%

\usetikzlibrary{matrix,arrows,automata,decorations.markings}

\newcommand{\eexists}{\ensuremath\ \exists\kern-.8em\exists\ }
\newcommand{\fforall}{\ensuremath\ \forall\kern-.8em\forall\ }

\newcommand{\tr}[2]{ \overset{#1}{\To}_{#2}}

\newtheorem{theorem}{Theorem}

\newtheorem{example}[theorem]{Example}

\renewcommand{\epsilon}{\varepsilon}

\newcommand{\C}{\mathcal{C}}
\newcommand{\V}{\mathcal{V}}
\newcommand{\qu}{\mathtt{Q}}

\newcommand{\To}{\longrightarrow}

\newcommand{\naturals}{\mathbb{N}}

\newcommand{\stefan}[1]{\textit{\textcolor{blue}{[stefan]: #1}}} 

\newcommand{\system}{\emph{WNetKAT}}

\begin{document}


\title{WNetKAT: A Weighted SDN Programming\\ and Verification Language}

%
%

\author{Kim G.~Larsen \quad Stefan Schmid \quad Bingtian Xue\\
{\small Aalborg University, Denmark}\\
{\texttt{\{kgl,schmiste,bingt\}}@cs.aau.dk}
}

\date{}

%



%
%

\maketitle

\sloppy

\begin{abstract}
Programmability and verifiability lie at the heart 
of the software-defined networking paradigm.
While OpenFlow and its match-action concept
provide primitive operations to manipulate hardware configurations,
over the last years, several more expressive
network programming languages have been developed.
This paper presents~$\system$, the first network
programming language accounting for the 
fact that networks are inherently weighted, and communications 
subject to capacity constraints (e.g., 
in terms of bandwidth) and costs
(e.g., latency or monetary costs). 
$\system$
is based on a syntactic and semantic
extension of the NetKAT 
algebra. We demonstrate several relevant
applications for~$\system$, including
cost- and capacity-aware reachability, 
as well as quality-of-service and fairness aspects.
These applications do not only apply
to classic, splittable and unsplittable~$(s,t)$-flows,
but also generalize to 
more complex network functions and
service chains. For example,
$\system$ allows to model flows which
need to traverse certain waypoint functions, which
may change the traffic rate.
This paper also shows the relation between the equivalence 
problem of WNetKAT and the equivalence problem of the weighted finite automata, which 
implies undecidability of the former. 
However, this paper also succeeds to 
prove the decidability of another useful problem,
which is sufficient in many practical scnearios: 
whether an expression equals to 0. 
Moreover, we initiate the discussion of 
decidable subsets of the whole language. 


\end{abstract}

\section{Introduction}

Managing and operating traditional computer networks is known
to be a challenging, manual and error-prone process. Given the
critical role computer networks play today, not only in the context
of the wide-area Internet but also of enterprise and data center networks,
this is worrisome. Software-Defined Networks (SDNs) in general
and the OpenFlow standard in particular, promise to
overcome these problems by
enabling automation, formal
reasoning and verification,
as well as 
by 
defining open standards for vendors. 
Indeed, there is also a wide consensus that formal verifiability is 
one of the key advantages of SDN over past attempts to
innovate computer networks, e.g., in the context of active networking~\cite{activenet}.
Accordingly, SDN/OpenFlow is seen as a promising paradigm
toward more dependable computer networks. 

At the core of the software-defined networking paradigm lies
the desire to program the network. 
In a nutshell, in an SDN, a general-purpose 
computer manages a set of programmable switches,
by installing rules (e.g., for forwarding) and reacting to events 
(e.g., 
newly arriving flows or link failures). 
In particular, OpenFlow follows a match-action paradigm:
the controller installs rules which define, using a 
\emph{match} pattern
(expressed over the \emph{packet header fields}, and defining a 
\emph{flow}),
which packets (of a flow) are subject to which \emph{actions} (e.g., forwarding
to a certain port).

While the OpenFlow API is simple and allows to manipulate
hardware configurations in flexible ways,
it is very low level 
and not well-suited as a language for 
human programmers.
Accordingly, over the last years, 
several more high-level and expressive domain-specific SDN languages
have been developed, especially within the Frenetic project~\cite{frenetic}. 
These languages can also be used to express fundamental network queries,
for example related to \emph{reachability}: 
They help administrators answer questions such as
\emph{``Can a given host~$A$ reach
host~$B$?''} or \emph{``Is traffic between hosts~$A$ and~$B$ 
isolated from traffic between hosts~$C$ and~$D$?''}.

What is missing today however is a domain-specific language
which allows to describe the important \emph{weighted aspects} of
networking. E.g., real networks naturally come with capacity constraints,
and especially in the Wide-Area Network (WAN) as well as in data centers,
bandwidth is a precious resource. Similarly, networks come with 
latency and/or monetary costs:
transmitting a packet over a wide-area link, or over a highly utilized link, may
entail a non-trivial latency, and inter-ISP links may also be attributed
with monetary costs.

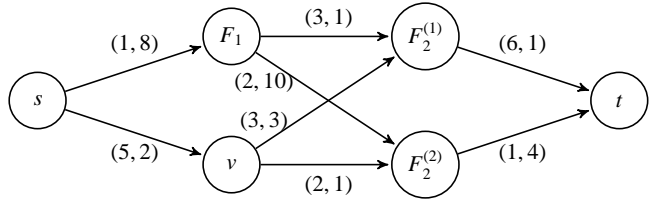
\begin{figure}[t]
\begin{center}
\begin{tikzpicture}[->,>=stealth',shorten >=1pt,auto,
                    semithick,scale=0.85, every node/.style={scale=0.85}]
  \tikzstyle{every state}=[fill=white,draw,text=black]
 \tikzstyle{vecArrow} = [thick, decoration={markings,mark=at position
   1 with {\arrow[semithick]{open triangle 60}}},
   double distance=1.4pt, shorten >= 5.5pt,
   preaction = {decorate},
   postaction = {draw,line width=1.4pt, white,shorten >= 4.5pt}]
\tikzstyle{innerWhite} = [semithick, white,line width=1.4pt, shorten >= 4.5pt]
    \node[state] (s) at (0,1) {$s$};   
    \node[state] (f1) at (3,2) {$F_1$};   
    \node[state] (e) at (3,0) {$v$};   
    \node[state] (f2-1) at (6,2) {$F_2^{(1)}$};   
    \node[state] (f2-2) at (6,0) {$F_2^{(2)}$};   
    \node[state] (t) at (9,1) {$t$};   
    
    \draw[->] (s) -- (f1);
    \node[above] at (1.5,1.6)  {$(1,8)$};
    
     \draw[->] (s) -- (e);
    \node[above] at (1.5,-0.1)  {$(5,2)$};

        \draw[->] (f1) -- (f2-1);
    \node[above] at (4.5,2)  {$(3,1)$};

                    \draw[->] (e) -- (f2-2);
    \node[above] at (4.5,-.6)  {$(2,1)$};

        \draw[->] (f2-1) -- (t);
    \node[above] at (7.5,1.6)  {$(6,1)$};

                        \draw[->] (f2-2) -- (t);
    \node[above] at (7.5,-0.1)  {$(1,4)$};

            \draw[->] (f1) -- (f2-2);
    \node[above] at (3.5,1)  {$(2,10)$};

                           \draw[->] (e) -- (f2-1);
    \node[above] at (3.5,0.4)  {$(3,3)$};
    
\end{tikzpicture}
\end{center}
\vspace{-.8em}
\caption{\emph{Example:} A network hosting two (virtualized) functions 
$F_1$ and~$F_2$. Function~$F_2$ is allocated twice. 
The functions~$F_1$ and~$F_2$ may 
change the traffic rate.}
\vspace{-1.3em}
\label{fig:sc}
\end{figure}

Weights may not be limited to links only, but also nodes (switches or routers)
have capacities and costs e.g., related to the 
packet rate. What is more, today's computer networks
provide a wide spectrum of in-network functions
related to security (e.g., firewalls) and performance
(e.g., caches, WAN optimizers).
To give an example, today, the number of 
so-called \emph{middleboxes} in enterprise networks
can be in the same order of magnitude as the number
of routers~\cite{someone}. 
A domain specific language for SDNs
should be expressive enough to account for middleboxes which can change
(e.g., compress or increase) the 
rate of the traffic passing through them.
Moreover, a network language should be able to 
define
that traffic must pass through these middleboxes  in the first place, i.e.,
that routing policies fulfill waypointing invariants~\cite{merlin}.
With the advent of more virtualized middleboxes, 
and the \emph{Network Function
Virtualization} paradigm, short \emph{NFV}, 
(virtualized) middleboxes may 
also be \emph{composed} to form more complex network services.
For example, SDN traffic engineering flexibilities
can be used to steer traffic through a series of middleboxes,
concatenating the individual functions into 
so-called \emph{service chains}~\cite{servicechains,sirocco15}.
For instance, a network
operator might want to ensure that all traffic from 
$s$ to~$t$ should first be routed
through a firewall~$FW$, and then through a WAN optimizer
$WO$, before eventually reaching~$t$:
the operator can do so by defining a service chain
$(s,FW,WO,t)$.
   

\noindent \textbf{A Motivating Example.}
Let us consider a more detailed example,
see the network in Figure~\ref{fig:sc}:
The network hosts two types of (virtualized) functions 
$F_1$ 
and~$F_2$: possible network functions may include, e.g., a firewall, 
a NAT, a proxy, a tunnel endpoint, a 
WAN optimizer (and its counterpart), a header decompressor, etc. 
In this example, function~$F_2$ is instantiated at two locations. 
Functions~$F_1$ and~$F_2$ may not be flow-preserving,
but may \emph{decrease} the traffic rate (e.g., in case of a proxy,
WAN optimizer, etc.) or \emph{increase} it: e.g., a tunnel entry-point
may add an extra header, a security box may add a
watermark to the packet, the counterpart of the WAN
optimizer may decompress the packet, etc. 
Links come with a certain cost (say latency)
and a certain capacity (in terms of bandwidth). Accordingly,
we may annotate links with two weights: the tuple~$(2,3)$ denotes that the
link 
cost is 2 and the link capacity 3.
We would like to be able to ask questions 
such as: \emph{Can source~$s$ emit traffic into the service chain
at rate~$x$ without overloading the network?}
or  \emph{Can we embed a service chain of cost (e.g., 
end-to-end latency) at most~$x$?}.
 
\noindent \textbf{Contributions.}
This paper initiates the study of weighted network 
languages
for programming
and reasoning about SDN networks, 
   which go beyond topological aspects but account
for actual resource availabilities, capacities, 
or costs. 
In particular, we present~$\system$, an extension of the \emph{NetKAT}~\cite{netkat}
algebra. 

For example,~$\system$ supports a natural generalization of 
the reachibility concepts used in classic network programming
languages, such as  
\emph{cost-aware} or \emph{capacity-aware reachability}. 
In particular, ~$\system$ allows to answer questions
of the form: \emph{Can host~$A$ reach host~$B$ at cost/bandwidth/latency~$x$?} 

We demonstrate applications of~$\system$ for a number
of practical use cases related to performance, quality-of-service, fairness, and costs.
These applications are not only useful in the context of both splittable
and unsplittable routing models, where flows need to travel from a source
$s$ to a destination~$t$, but also in the context of more complex
models with waypointing requirements (e.g., 
service chains).

The weighted extension of NetKAT is non-trivial,
as capacity constraints introduce dependencies 
between flows, and 
arithmetic operations such as \emph{addition} (e.g., in case of latency) 
or \emph{minimum} (e.g., in case of bandwidth to compute the end-to-end delay) have to be supported along the
paths.
Therefore, we extend the syntax of NetKAT 
toward weighted packet- and switch-variables, as well as queues, 
and provide a semantics accordingly.
In particular, one contribution of our work is to show
for which weighted aspects and use cases which language extensions are required.

We also show the relation between WNetKAT expressions 
and weighted finite automata~\cite{droste2009handbook} -- 
an important operational model for weighted programs. 
This leads to the undeciability of WNetKAT equivalence problem. 
However,
leveraging this relation we also succeed to 
prove the decidability of whether an expression equals to 0: 
for many practical scenarios a sufficient and relevant
solution. Moreover, this paper initiates the discussion of identifying 
decidable subsets of the whole language. 



\noindent \textbf{Related Work.}
Most modern domain-specific SDN languages 
enable automated tools for 
verifying network properties~\cite{pane,frenetic,pyretic,nettle,maple}. 
Especially reachability
properties, which are also the focus in our paper, 
have been studied intensively in the literature~\cite{hsa,veriflow}.
Indeed, the formal verifiability
of the OpenFlow 
match-action interface~\cite{hsa,veriflow,padon2015decentralizing,zhou15ccg}
constitutes a key advantage of the paradigm over
previous innovation efforts~\cite{active-nets}. 
Existing expressive languages use SAT formulas~\cite{anteater}, 
graph-based representations~\cite{hsa,veriflow}, or higher-order
logic~\cite{fvf} to describe network topologies and policies. 

Our work builds upon NetKAT, a new framework 
based on Kleene algebra with
tests
for specifying, programming,
and reasoning about networks and policies. 
NetKAT respresents a more principled approach compared
to prior work, and is also motivated by the observation that
end-to-end functionality is determined not only by the behavior of the
switches and but also by the structure of the network topology. 
NetKAT in turn is
based on earlier efforts performed in the context of NetCore~\cite{netcore}, 
Pyretic~\cite{pyretic} and Frenetic~\cite{frenetic}. It has recently been extended
to a probabilistic setting~\cite{pnetkat}.
The Kleene algebra with tests was developed by
 Kozen~\cite{kat-dexter}. 
 
However, to the best of our knowledge, there is 
prior work on weighted versions of NetKAT.
 

\noindent \textbf{Organization.}
The rest of this paper is organized as follows.
Section~\ref{sec:background} provides the necessary background
on SDN and NetKAT.
Section~\ref{sec:wnetkat} introduces~$\system$, our weighted version of NetKAT.
Section~\ref{sec:applications} demonstrates the usefulness of our extensions
in a number of applications.
Section~\ref{sec:deci} 
and Section~\ref{sec:compilation}  discuss 
complexity and implementation aspects.
Section~\ref{sec:conclusion} concludes our contribution.


\section{Background}\label{sec:background}

We first provide a more detailed introduction to OpenFlow, 
and then describe the programming language NetKAT, 
which compiles to OpenFlow.

\noindent \textbf{SDN and OpenFlow.}
A Software-Defined Network (SDN) outsources and consolidates
the control over 
data plane elements to a logically
centralized control plane implemented in software.
%
Arguably, software-defined networking in general,
and its de facto standard, OpenFlow, 
are 
about programmability,
verifiability and generality~\cite{road}:
 A software-defined network 
 allows programmers to write network applications (for example
for traffic engineering) in software. 
The behavior of an OpenFlow switch is defined
by its configuration: a list of
prioritized \emph{(flow) rules} 
stored in the switch flow table,
which are used to classify, filter, and modify packets
based on their \emph{header fields}.
In particular, OpenFlow follows a simple match-action paradigm:
 the match parts of
the flow
rules  (expressed over the header fields) 
specify which packets belong to a certain flow (e.g., 
depending on the IP destination address), and the action parts
define how these packets should be processed (e.g., forward 
to a certain port).
OpenFlow supports a rather general
packet processing: it allows to match and process 
packets based on their Layer-2 (e.g., MAC addresses),
Layer-3 (e.g., IP addresses), and Layer-4 header fields
(e.g., TCP ports), 
or even in a protocol-independent manner, using arbitrary bitmasking~\cite{p4}. 
For example,
an OpenFlow router may forward packets destined to http ports differently from traffic destined to
ftp ports.
In other words, an OpenFlow switch
blurs the difference between
switches and routers (the two terms are used interchangeably in this paper), and even supports some basic middlebox functionality.

OpenFlow also readily supports quantitative
aspects, e.g., the selection of queues annotated with different round robin
weights (the standard approach to implement quality-of-service guarantees in networks today), or meters (measuring
the bandwidth of a flow). 
Moreover, we currently
witness a trend toward more flexible
and stateful programmable
switches and packet processors, 
featuring group tables, counters, 
and beyond~\cite{brain,p4,snap,openstate,pof,domino,ovs}.

\noindent \textbf{NetKAT.}
The formal framework developed in this paper
is based on NetKAT~\cite{netkat}.
Here we briefly review the main
concepts underlying NetKAT, and discuss how they relate to OpenFlow.

NetKAT is a
high-level
algebra for reasoning about network programs. It is
based on \emph{Kleene Algebra with Tests (KAT)}, 
and uses an equational theory combining the axioms of KAT
and network-specific axioms that describe transformations on packets
(as performed by OpenFlow switch rules).
These axioms facilitate reasoning about local switch processing
functionality (needed during compilation and 
for optimization) as well as
global network behavior (needed to check reachability and traffic
isolation properties). Basically, an atomic NetKAT policy 
(a function from packet headers
to sets of packet headers: essentially the per-switch
OpenFlow rules discussed 
above) 
can be used to filter or modify packets. 
Policy combinators ($+$) allow to build larger policies out
of smaller policies.
There is also a sequential composition combinator 
to apply functions consecutively.

Besides the \emph{policy}, modeling the per-switch OpenFlow rules, 
a network programming language needs
to be able to describe the network \emph{topology}. 
NetKAT models the network topology as a directed graph:
nodes (hosts, routers, switches) are connected via
edges (links) using (switch) \emph{ports}.
NetKAT simply describes
the topology
as the union of smaller policies that encode the behavior of each
link. To model the effect of sending
a packet across a link,
NetKAT employs the sequential composition
of a filter that retains packets located at one end of the link, and
a modification that updates the switch and port fields to the location at
the other end of the link. 
Note that the NetKAT topology and the 
NetKAT policy are hence to be seen as two independent concepts.
Succinctly: 

A \emph{Kleene algebra} (KA) is any structure~$(K,+,\cdot,^\ast,0,1)$, where~$K$ is a set,~$+$ and~$\cdot$ are binary operations on~$K$,~$^\ast$ is a unary operation on~$K$, and~$0$ and~$1$ are constants, satisfying the following axioms, 
where we define~$p\leq q$ iff~$p+q=q$.
\\\centerline{$\begin{array}{lcl}
p+(q+r)=(p+q)+r&\phantom{egef}&p(qr)=(pq)r\\
p+q=q+p&&1\cdot p=p\cdot 1=p\\
p+0=p+p=p&&p\cdot 0=0\cdot p=0\\
p(q+r)=pq+pr&&(p+q)r=pr+qr\\
1+pp^\ast\leq p^\ast&&q+px\leq x\Rightarrow p^\ast q\leq x\\
1+p^\ast p\leq p^\ast &&q+xp\leq x\Rightarrow qp^\ast\leq x
\end{array}$}

A \emph{Kleene algebra with tests} (KAT) is a two-sorted structure~$(K,B,+,\cdot,^\ast,\overline{\ },0,1)$, where~$B\subseteq K$ and 
\\$\bullet$ $(K,+,\cdot,^\ast,0,1)$ is a Kleene algebra;
\\$\bullet$ $(B,+,\cdot,\overline{\ },0,1)$ is a Boolean algebra;
\\$\bullet$ $(B,+,\cdot,0,1)$ is a subalgebra of~$(K,+,\cdot,0,1)$.

The elements of~$B$ are called \emph{tests}.
The axioms of Boolean algebra are:
\\\centerline{$\begin{array}{lcl}
a+bc=(a+b)(a+c)&\phantom{egef}&ab=ba\\
a+1=1&&a+\overline{a}=1\\
a\overline{a}=0&&aa=a
\end{array}$}

NetKAT is a version of KAT in which
  the atoms (elements in $K$) are 
defined over header fields~$f$ (variables)
and values~$\omega$:
\\$\bullet$ $f\leftarrow \omega$  \hfill (``assign a value~$\omega$ to header field~$f$'')
\\$\bullet$ $f=\omega$ \hfill (``test the value of a header field'')
\\$\bullet$ {\small{\sf dup}} \hfill (``duplicate the packet'')

The set of all possible values of $f$ is denoted $\Omega$. For readability, we use~$skip$ and~$drop$ to denote~$1$ and~$0$, respectively.

The NetKAT axioms consist of the following equations,
in addition to the KAT axioms on the commutativity and redundancy 
of different actions and tests, 
and enforcing that the field has exactly one value:
\\\centerline{$\begin{array}{rcl}
f_1\leftarrow \omega_1;f_2\leftarrow \omega_2&=&f_2\leftarrow \omega_2;f_1\leftarrow \omega_1\phantom{gg}(f_1\not=f_2)\phantom{eggg}\hfill(1)\\
f_1\leftarrow \omega_1;f_2=\omega_2&=&f_2=\omega_2;f_1\leftarrow \omega_1\phantom{gg}(f_1\not=f_2)\hfill(2)\\
f=\omega;\text{\small\sf dup}&=&\text{\small\sf dup};f=\omega\hfill(3)\\
f\leftarrow \omega;f=\omega&=&f\leftarrow \omega\hfill(4)\\
f=\omega;f\leftarrow \omega&=&f=\omega\hfill(5)\\
f\leftarrow \omega_1;f\leftarrow \omega_2&=&f\leftarrow \omega_2\hfill(6)\\
f=\omega_1;f=\omega_2&=&0 \phantom{ejggg}(\omega_1\not=\omega_2)\hfill(7)\\
\displaystyle\sum_{\omega\in \Omega}f=\omega&=&1\hfill(8)
\end{array}$}

In terms of semantics, NetKAT uses \emph{packet histories} to record the state of each
packet on its path from switch to switch through the
network. 
The notation 
$\langle pk_1, \ldots, pk_n \rangle$ 
is used to describe a history with
elements~$pk_1, \ldots, pk_n$ being packets;
$pk::\langle\rangle$ is used to denote a
history with one element and~$pk::h$ 
to denote the history constructed
by prepending~$pk$ on to~$h$. 
By convention, the first element of a history is
the current packet (the ``head'').
A NetKAT expression denotes
a function 
$\llbracket\ \rrbracket: H\rightarrow 2^H$, where~$H$ is the set of packet histories.
Histories are only needed for reasoning:
Policies only inspect or modify the first (current)
packet in the history. 
Succinctly:
\\\centerline{$\begin{array}{rcl}
\llbracket f\leftarrow \omega\rrbracket(pk::h)&\phantom{e}=\phantom{e}&\{pk[\omega/f]::h\}\\
\llbracket f=\omega\rrbracket(pk::h)&=&\left\{\begin{array}{lcl}
\{pk::h\}&\phantom{ef}&\text{if }pk(f)=\omega\\\emptyset&&\text{otherwise}
\end{array}\right.\\
\llbracket\text{\small\sf dup}\rrbracket(pk::h)&=&\{pk::pk::h\}\\
\llbracket p+q\rrbracket(h)&=&\llbracket p\rrbracket(h)\cup\llbracket q\rrbracket(h)\\
\llbracket pq\rrbracket(h)&=&\bigcup_{h'\in\llbracket p\rrbracket(h)}{\llbracket q\rrbracket(h')}\\
\llbracket p^\ast\rrbracket(h)&=&\bigcup_{n}{\llbracket p^n\rrbracket(h)}\\
\llbracket 0\rrbracket(h)&=&\emptyset\\
\llbracket 1\rrbracket(h)&=&\{h\}\\
\llbracket\overline{a}\rrbracket(h)&=&\left\{\begin{array}{lcl}
\{h\}&\phantom{ef}&\text{if }\llbracket a\rrbracket(h)=\emptyset\\\emptyset&&\text{if }\llbracket a\rrbracket(h)=\{h\}
\end{array}\right.
\end{array}$}

\begin{example}
Consider the network in Figure~\ref{fig:sc}. 
NetKAT can be used
to specify the topology as follows, where the field~$sw$ stores 
the current location (switch) of the packet:
\\\centerline{$\begin{array}{lcl}
t&::=&\ sw=s;(sw\leftarrow F_1+sw\leftarrow v)\\
&&+sw=F_1;(sw\leftarrow F_2^{(1)}+sw\leftarrow F_2^{(2)})\\
&&+sw=v;(sw\leftarrow F_1^{(1)}+sw\leftarrow F_2^{(2)})\\
&&+sw=F_2^{(1)};sw\leftarrow t\\
&&+sw=F_2^{(2)};sw\leftarrow t
\end{array}
$}

\noindent The first line of the above NetKAT expression specifies that if the packet 
is at~$s$, then it will be sent to~$F_1$ or~$v$. 
Analogously for the other cases.
In OpenFlow, this policy can be implemented using OpenFlow rules,
whose match part applies to packets arriving at $s$, 
and whose action part assigns the packets to the respective
forwarding ports.\hfill~$\blacksquare$  
\end{example}

However, one can observe that with NetKAT it is not possible to specify or reason about the important
quantitative aspects in Figure~\ref{fig:sc}, e.g., the cost and capacity along 
the links or the function of $F_2$ which changes the rate of the flow.  To do these, a weighted extension of NetKAT is needed. 

\section{WNetKAT}\label{sec:wnetkat}

On a high level, a computer network can be described as a set of
nodes (hosts or routers) which are interconnected by a set of
links, hence defining the network topology. While this high-level view is sufficient for many purposes,
for example for reasoning about reachability, in practice, the situation is 
often more complex: both nodes and links come with 
capacity constraints (e.g., in
terms of buffers, CPU, and bandwidth) and may be attributed with costs (e.g., monetary
or in terms of performance). In order to reason about performance, cost,
and fairness aspects,
it is therefore important to take these dimensions into account. 

The challenge
of extending NetKAT to weighted scenarios
lies
in the fact that in a weighted 
network, traffic flows can no longer be considered independently,
but they may \emph{interfere}: their packets compete for 
the shared resource. 
Moreover, packets of a given flow may not necessarily
be propagated along a unique path, but may be
split and distributed among multiple paths (in the so-called
\emph{multi-path routing} or \emph{splittable flow} variant). Accordingly,
a weighted extension of NetKAT must be able to
deal with ``inter-packet states''.
 
We in this paper will think of the network as a 
weighted (directed) graph~$G=(V,E,w)$.
Here,~$V$ denotes the set of switches (or equivalently routers, and
henceforth often simply called nodes),~$E$
is the set of links (connected to the switches by \emph{ports}), and
$w$ is a weight function. The weight function~$w$ applies 
to both nodes~$V$ as well as links~$E$. Moreover, a node
and a link may be characterized by \emph{a vector of weights}
and also combine \emph{multiple resources}:
for example, a list of capacities (e.g., CPU and memory
on nodes, or bandwidth on links) and a list of costs
(e.g., performance, energy, or monetary costs).

In order to specify the quantitative aspects, 
we propose in this paper a weighted extension of NetKAT: $\system$. 
In addition to NetKAT:
\begin{itemize}
\item $\system$ includes a set of \emph{quantitative packet-variables}
to specify the 
quantitative information carried in the packet, in addition
to the regular
(non-quantitative) packet-variables of NetKAT
(called \emph{fields} in NetKAT): e.g., regular variables are used to 
describe locations, such as switch
and port, or priorities, while quantitative variables are used to specify latency or energy. The set of all packet-variables is denoted by $\V_p$. 
\item $\system$ also includes a set of \emph{switch-variables}, denoted by~$\V_s$, 
to specify the configurations at the switch. 
Switch variables can either be quantitative (e.g., counters, meters, 
meta-rules~\cite{p4,ccr16sync})
or non-quantitative (e.g., location related), as it is
the case of the packet-variables.
\end{itemize}

\noindent\emph{Remarks:} The set of quantitative (packet- and switch-) variables is denoted by $\V_q$ and these variables
range over the natural numbers~$\naturals$ 
(e.g., normalized rational numbers). 
The set of non-quantitative (packet- and switch-) variables is denoted $\V_n$ and the set of the possible values is denoted
$\Omega$. 
Note that~$\V_q\cap\V_n=\emptyset$ and~$\V_q\cup\V_n=\V_p\cup\V_s$.\hfill~$\blacksquare$ 

In addition to introducing quantitative variables, we also 
need to extend the atomic actions and tests of NetKAT. 
Concretely,~$\system$ first supports 
non-quantitative assignments 
and non-quantitative tests on the non-quantitative switch-variables, 
similar to those on the packet-variables in NetKAT. Moreover, 
$\system$ also allows for
\textit{quantitative assignments} and \emph{quantitative tests}, 
defined as follows, where~$x\in\V_q$, 
$\V'\subseteq\V_q$,~$\delta\in\naturals$,~$\bowtie\in\{>,<,\leq,\geq,=\}$:
\begin{itemize}
\item  \textbf{Quantitative Assignment}~$x\leftarrow (\Sigma_{x'\in\V'}{x'}+\delta)$: 
Read the current values of the variables in~$\V'$ and add them 
to~$\delta$, then assign this result to~$x$.
\item \textbf{Quantitative Test}~$x\bowtie (\Sigma_{x'\in\V'}{x'}+\delta)$: Read the current 
value of the variables in~$\V'$ and add them to~$\delta$, 
then compare this result to the current value of~$x$.
\end{itemize} 

\noindent\emph{Remarks:} 1. In the quantitative assignment and test, 
only addition is allowed. However, 
an extension to other arithmetic operations 
(e.g., linear combinations)
is straightfoward. Moreover, calculating~\emph{minimum} 
or~\emph{maximum} may be useful in practice: e.g., the throughput of a flow often depends on the weakest link (of minimal bandwidth) along a path. 
Note that these operations can actually be 
implemented with quantitative assignments and tests, i.e., 
by comparing 
every variable 
to another and determining the smallest. E.g.,
for $x\in\V_q$ and~$y,z\in\V_q$ or~$\naturals$,
\\\centerline{$x\leftarrow\min\{y,z\}\overset{\textsf{def}}{=}\ 
y\leq z;x\leftarrow y\ \&\ y>z;x\leftarrow z$} \\
2. In quantitative assignment and test, $x$ might be in $\V'$.\\
3. We use~$+$ to denote the arithmetic operation over numbers. Therefore, we will use ``$\&$" in~$\system$ to denote the ``$+$" operator of Kleene Algebra, which is also used in~\cite{pnetkat}.
\hfill~$\blacksquare$ 


Given the set of switches $V$, a \emph{switch-variable valuation} is a partial function
$\rho:V\times \V_s\hookrightarrow\naturals\cup\Omega$.
It associates, for each switch and each switch-variable, 
a integer or a value from~$\Omega$. 
We emphasize that~$\rho$ is a 
partial function, as some variables may not be defined at some switches.

A \system expression denotes a function
$\llbracket\ \rrbracket: \rho\times H\rightarrow 2^{H}$, where~$H$ is 
the set of packet histories. The semantics of~$\system$ is defined in Table~\ref{sem-1}, where $x\in\V_n,y\in\V_q$,~$\delta\in\naturals$ and~$\omega\in\Omega$.

\begin{table*}[]
$$\begin{array}{rcl}
\llbracket x\leftarrow \omega\rrbracket(\rho,\ pk::h)&
\phantom{e}=\phantom{e}&
\left\{\begin{array}{lcl}
\{\rho,\ pk[\omega/x]::h\}&\phantom{ef}&\text{if }x\in\V_p
\\
\{\rho(v)[\omega/x],\ pk::h\}&&\text{if }x\in\V_s\text{ and }pk(sw)=v
\\
\end{array}
\right.
 \hfill (1)
 \\
 \\
\llbracket x=\omega\rrbracket(\rho,\ pk::h)&=&\left\{\begin{array}{lcl}
\{\rho,\ pk::h\}&\phantom{ef}&\text{if }x\in\V_p\text{ and }pk(x)=\omega
\\
&\phantom{ef}&\text{ or if }x\in\V_s,pk(sw)=v\text{ and }\rho(v,x)=\omega
\\
\emptyset&&\text{otherwise}
\end{array}\right.
\hfill (2)
\\
\\
\llbracket y\leftarrow (\displaystyle\Sigma_{y'\in\V'}{y'}+r)\rrbracket(\rho,\ pk::h)&\phantom{e}=\phantom{e}&\left\{\begin{array}{lcl}
\{\rho,\ pk[r'/x]::h\}&\phantom{ef}&\text{if }x\in\V_p
\\
\{\rho(v)[r'/x],\ pk::h\}&&\text{if }x\in\V_s\text{ and }pk(sw)=v
\\
\end{array}
\right.\hfill (3)\\
&\multicolumn{2}{l}{\text{where }r'=\Sigma_{y_p\in\V'\cap\V_p}pk(y_p)+\Sigma_{y_s\in\V'\cap\V_q}\rho(v,y_s)+r}
\\
\\
\llbracket y=(\Sigma_{y'\in\V'}{y'}+r)\rrbracket(\rho,\ pk::h)&=&\left\{\begin{array}{lcl}
\{\rho,\ pk::h\}&\phantom{ef}&\text{if }x\in\V_p\text{ and }pk(x)=r'\\
&\phantom{ef}&\text{ or }x\in\V_s,pk(sw)=v\text{ and }\rho(v,x)=r'\\
\emptyset&&\text{otherwise}
\end{array}\right.
\hfill (4)\\
&\multicolumn{2}{l}{\text{where }r'=\Sigma_{y_p\in\V'\cap\V_p}pk(y_p)+\Sigma_{y_s\in\V'\cap\V_q}\rho(v,y_s)+r}
\end{array}$$
\caption{Semantics of~$\system$}\label{sem-1}
\vspace{-1.8em}
\end{table*}

\noindent\emph{Remarks:}  $\bullet$ Equations (1) and (3) update the corresponding header field if~$x$ is a packet-variable, or they update the corresponding switch information of the current switch if~$x$ is a switch-variable. 
Equation~(1) updates the non-quantitative variables and Equation~(3) the quantitative ones.
$\bullet$ Equations (2) and (4) test the non-quantitative and quantitative variables respectively, using the current packet- and switch-variables.\hfill~$\blacksquare$

\begin{example}\label{exm-sem}
Consider again the network in Figure~\ref{fig:sc}. The topology of the network can be characterized with the following~$\system$ formula~$t$, where~$sw$ specifies the current location (switch) of the packet,~$co$ specifies the cost, and~$ca$ specifies the capacity along the links. 

\centerline{$\begin{array}{ll}
t::=&sw=s;(sw\leftarrow F_1;co\leftarrow co+1;ca\leftarrow\min\{ca,8\} \\
&\phantom{sw=s;}\&\ sw\leftarrow v;co\leftarrow co+5;ca\leftarrow\min\{ca,2\})\\
&\&\ sw=F_1;\\
&\phantom{sw=;}(sw\leftarrow F_2^{(1)};co\leftarrow co+3;ca\leftarrow\min\{ca,1\}\\
&\phantom{sw=}\&\ sw\leftarrow F_2^{(2)};co\leftarrow co+2;ca\leftarrow\min\{ca,10\})\\
&\&\ sw=v;(sw\leftarrow F_2^{(1)};co\leftarrow co+3;ca\leftarrow\min\{ca,3\}\\
&\phantom{sw=f}\&\ sw\leftarrow F_2^{(2)};co\leftarrow co+2;ca\leftarrow\min\{ca,1\})\\
&\&\ sw=F_2^{(1)};sw\leftarrow t;co\leftarrow co+6;ca\leftarrow\min\{ca,1\}\\
&\&\ sw=F_2^{(2)};sw\leftarrow t;co\leftarrow co+1;ca\leftarrow\min\{ca,4\}
\end{array}
$}

The variable~$co$ accumulates the costs along the path, 
and the variable~$ca$ records the smallest capacity along the path.
Notice that~$ca$ is just a packet-variable used to record the capacity of the path; 
it does not represent the capacity used by this packet (the latter is 
assumed to be negligible). 

Assume that function~$F_1$ is flow conserving
(e.g., a NAT),
while $F_2$ increases the flow rate
by an additive constant $\gamma\in\naturals$ (e.g., a security related
function, adding a watermark or an IPSec header).
The policy of $F_2$ can be specified as:
\centerline{$\begin{array}{ll}
p_{F_2}::=&(sw=F_2^{(1)}\ \&\ sw=F_2^{(2)});ca\leftarrow ca+\gamma
\end{array}
$\hfill~$\blacksquare$ }
\end{example}

\noindent\emph{Remarks:} Note that this simple example required only
(non-quantitative and quantitative) packet-variables. 
However, as we will see in Section \ref{sec:applications}, 
to model  
more complex aspects of networking, such as splittable flows, 
additonal concepts of $\system$ will be needed. 
\hfill~$\blacksquare$


\section{Applications}\label{sec:applications}

The weighted extensions introduced by~$\system$
come with a number of interesting applications. In this section,
we show that the notions of reachability frequently
discussed in prior work, find natural extensions
in the world of weighted networks, and discuss applications in the context of
service chains, fairness, and quality-of-service.
In the Appendix, additional details are provided for some of these use cases.


\subsection{Cost Reachability} \label{sec:app-cost}

Especially data center networks but also wide-area networks,
and to some extent enterprise networks, feature a certain
\emph{path diversity}~\cite{path-diversity}: there exist multiple routes between two
endpoints (e.g., hosts). This path diversity is not only 
a prerequisite
for fault-tolerance, but also introduces traffic engineering flexibilities.
In particular, different paths or routes depend on different
links, whose cost can vary. 
For example, links may be attributed
with 
monetary costs: a peering link may be free of charge,
while an up- or down-link is not. 
Links cost can also be performance related, and may for
example vary in terms of latency, for example due to the
use of different technologies~\cite{speed-of-light}, or simply
because of different physical distances. The monetary and
performance costs are often related: for example, in the 
context of stock markets, lower latency links come at a
higher price~\cite{nordunet}.
It is therefore natural to ask questions such as:
\emph{``Can A reach B at cost at most~$c$?''}.
We will refer to this type of questions as
\emph{cost reachability questions}.

\begin{example}\label{exm-cr} Consider the network in Figure~\ref{fig:b4}. The topology roughly describes
the North American data centers interconnected by Google B4, according
to~\cite{b4}.
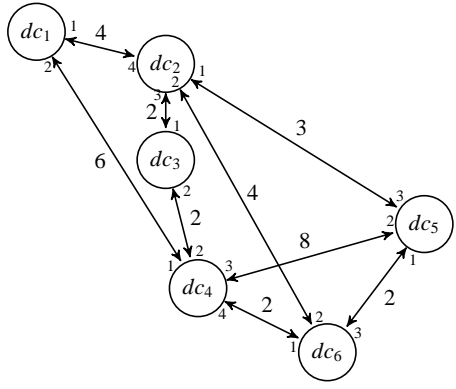
\begin{figure}[!htb]
\vspace{-.8em}
\begin{center}
\begin{tikzpicture}[->,>=stealth',shorten >=1pt,auto,
                    semithick,scale=0.85, every node/.style={scale=0.85}]
  \tikzstyle{every state}=[fill=white,draw,text=black]
 \tikzstyle{vecArrow} = [thick, decoration={markings,mark=at position
   1 with {\arrow[semithick]{open triangle 60}}},
   double distance=1.4pt, shorten >= 5.5pt,
   preaction = {decorate},
   postaction = {draw,line width=1.4pt, white,shorten >= 4.5pt}]
\tikzstyle{innerWhite} = [semithick, white,line width=1.4pt, shorten >= 4.5pt]
    \node[state] (dc1) at (0,0) {$dc_1$};   
    \node  at (0.55,0.1) {\scriptsize~$1$}; 
    \node  at (0.15,-0.55) {\scriptsize~$2$}; 
    
    \node[state] (dc2) at (2,-.5) {$dc_2$};   
    \node  at (2.54,-0.6) {\scriptsize~$1$}; 
    \node  at (2.12,-0.78) {\scriptsize~$2$}; 
    \node  at (1.85,-0.98) {\scriptsize~$3$}; 
    \node  at (1.45,-0.55) {\scriptsize~$4$};

    \node[state] (dc3) at (2,-2) {$dc_3$};   
    \node  at (2.3,-2.48) {\scriptsize~$2$}; 
    \node  at (2.15,-1.47) {\scriptsize~$1$}; 
    
    \node[state] (dc4) at (2.5,-4) {$dc_4$};   
    \node  at (2.05,-3.65) {\scriptsize~$1$}; 
    \node  at (2.5,-3.42) {\scriptsize~$2$}; 
    \node  at (2.95,-3.65) {\scriptsize~$3$}; 
    \node  at (2.85,-4.4) {\scriptsize~$4$}; 
    
    \node[state] (dc5) at (6,-3) {$dc_5$};   
    \node  at (5.8,-3.55) {\scriptsize~$1$}; 
    \node  at (5.45,-2.95) {\scriptsize~$2$};
    \node  at (5.6,-2.56) {\scriptsize~$3$};  
    
    \node[state] (dc6) at (4.5,-5) {$dc_6$};   
    \node  at (3.95,-4.92) {\scriptsize~$1$}; 
    \node  at (4.35,-4.45) {\scriptsize~$2$}; 
    \node  at (4.95,-4.7) {\scriptsize~$3$}; 
    
    \draw[<->] (dc1) -- (dc2);
    \node[above] at (1,-0.25) {$4$};
    
    \draw[<->] (dc2) -- (dc3);
    \node[left] at (2,-1.25) {$2$};
    
    \draw[<->] (dc3) -- (dc4);
    \node[right] at (2.25,-2.9) {$2$};
    
    \draw[<->] (dc1) -- (dc4);
    \node[left] at (1.2,-2) {$6$};
    
    \draw[<->] (dc2) -- (dc5);
    \node[above] at (4.1,-1.75) {$3$};
    
    \draw[<->] (dc2) -- (dc6);
    \node[above] at (3.35,-2.75) {$4$};
    
    \draw[<->] (dc4) -- (dc5);
    \node[above] at (4.15,-3.5) {$8$};
    
    \draw[<->] (dc4) -- (dc6);
    \node[above] at (3.57,-4.5) {$2$};
    
    \draw[<->] (dc5) -- (dc6);
    \node[right] at (5.25,-4.15) {$2$};
\end{tikzpicture}
\end{center}
\vspace{-.8em}
\caption{Example topology: excerpt of Google B4~\cite{b4} (U.S.~data centers only).
Nodes here represent data centers (resp.~OpenFlow switches located 
at the end of the corresponding long-haul fibers). 
Links are annotated with weights, and nodes are interconnected 
via ports (\emph{small numbers}).
} 
\label{fig:b4}
\vspace{-.5em}
\end{figure}

In order to reason about network latencies, we 
not only need information about the 
switch at which the packet is currently located 
(as in our earlier 
examples), but also the \emph{port of the switch} needs to be specified. 
We introduce the packet-variable~$pt$.
We can then specify this network topology in~$\system$.
The link from~$dc_1$ to~$dc_2$ 
(latency~$4$ units)
represented by 
the port~$1$ at 
$dc_1$ and the port~$4$ at~$dc_2$ is specified as follows, where we use packet-variable~$sw$ to denote the current switch, 
$pt$ to specify the current port, and~$l$ to specify 
the latency of the path the packet traverses,

\centerline{$sw=dc_1;pt=1;sw\leftarrow dc_2;pt\leftarrow 4;l\leftarrow l+4$}
Analogously, the entire network topology can be modeled with~$\system$, 
henceforth denoted by~$t$. 
The policy of the network determines the functionality
of each switch (the OpenFlow rules), 
e.g., in~$dc_2$, packets from $dc_1$ to $dc_5$ arriving at port~$4$ are always sent out through 
port~$1$ or port~$3$. This can be specified as:
\\\centerline{$src=dc_1;dst=dc_5;sw=dc_2;pt=4;(pt\leftarrow 1\&\ pt\leftarrow 3)$}
Analogously, the entire network policy can be modeled with~$\system$, 
henceforth denoted by~$p$. 

To answer the cost reachability question,
one can check whether the following~$\system$ expression is equal to~$drop$. 
\\\centerline{$scr\leftarrow A;dst\leftarrow B;l\leftarrow 0;sw\leftarrow X;pt(pt)^\ast;sw=B;l\leq c$} 

If it is equal to~$drop$, then~$B$ cannot be reached from~$A$ at latency at most~$c$; otherwise, it can.\hfill ~$\blacksquare$ 
\end{example}

\noindent\emph{Remarks:} For ease of presentation, in the above example, 
we considered only one weight. 
However, $\system$ readily supports
multiple weights: we 
can simply use multiple variables accordingly. 
Moreover, while the computational problem complexity
can increase with the number of considered weights~\cite{korkmaz2001multi}, 
the multi-constrained path selection does not affect the general 
asymptotic complexity of $\system$. \hfill ~$\blacksquare$

\subsection{Capacitated Reachability}\label{sec:app-cap}

Especially in the wide-area network, but also in data centers,
link capacities are a scarce resource: indeed, wide-area traffic
is one of the fastest growing traffic aggregates~\cite{b4}. 
However, also the routers themselves come with 
capacity constraints, both in terms of memory (size of TCAM)
as well as CPU: for example, the CPU utilization has been shown
to depend on the packet rate~\cite{packet-rate}.
Accordingly, a natural question to ask is: \emph{Can 
A communicate at rate at least~$r$ to B?}
We will refer to this type of questions as
\emph{capacitated reachability questions}. 

There are two problem variants: 

\begin{itemize}
\item \emph{Unsplittable flows:} The capacity needs
to be computed along a single path (e.g., an MPLS tunnel). 
\item \emph{Splittable flows:} The capacity
needs to be computed along multiple paths 
(e.g., MPTCP, ECMP). 
We will assume links of higher capacity
are chosen first.
\end{itemize}

For both variants, to find out the capacity of paths between two nodes, a \emph{single} test packet will be sent to explore the network and record the bandwidth/capacity with a packet-variable in the packet. 
We assume that the bandwidth consumed by this packet is negligibile. 
Also, only once the packet has traversed and determined the bandwidth, 
e.g., the actual (large) flows are allocated accordingly (by the SDN controller).

\begin{example}
Consider the network in Figure~\ref{fig:b4} again, but assume that the labels are the capacities rather than latency. 

\noindent\textbf{Unsplittable flow scenario:}
The switch policies are exactly the same as 
in Example~\ref{exm-cr}, while the topology will be specified similarly 
using packet-variable~$c$ to record the capacity of the link. 
E.g., the link between~$dc_1$ and~$dc_2$ can be specified as:
\\\centerline{$sw=dc_1;pt=1;sw\leftarrow dc_2;pt\leftarrow 4;
c\leftarrow \min\{c,4\}$}

The unsplittable capacitated reachability question can be answered by checking whether the following expression is equal to~$drop$, 
\\\centerline{$scr\leftarrow A;dst\leftarrow B;c\leftarrow r;sw\leftarrow A;pt(pt)^\ast;sw=B;c\geq r$}

If the above formula does not equal~$drop$, then~$A$ can communicate at rate at least~$r$ to~$B$.

Another (possibly) more efficient approach is not to update~$c$ 
while the bandwidth is smaller than~$r$ (meaning that a flow of 
size~$r$ cannot go through this link). In this case, one can specify the topology 
as follows, where~$c$ is not used to record the capacity along the path anymore, 
but rather to test whether this link is wide enough:
\\\centerline{$sw=dc_1;pt=1;sw\leftarrow dc_2;pt\leftarrow 4;c\leq 4$}

The above~$\system$ expression only tests 
whether~$c$ is less than or equal to~$4$. 
It makes sure that the value of~$c$ (which is~$r$) 
does not exceed the capacity of the following link. 
If it exceeds the capacity of the link, then a flow of 
rate~$r$ cannot use this link. Therefore, the test packet is 
dropped already. 
The capacitated reachability question can then be 
answered by checking whether the following expression is equal to~$drop$:
\\\centerline{$scr\leftarrow A;dst\leftarrow B;c\leftarrow r;sw\leftarrow A;pt(pt)^\ast;sw=B$}

If the above formula does not equal~$drop$, then~$A$ can 
communicate at rate at least~$r$ to~$B$.

\noindent\textbf{Splittable flow scenario:} 
For the splittable scenario, the situation is far more complicated. 
For example, in~$dc_2$, packets arriving 
at port~$4$ are sent out through port~$1$ or port~$2$, 
and port~$2$ prioritizes port~$1$. That is, if the incoming traffic
has rate~$4$, then a share of $3$ units will be sent out through 
port~$2$, and a~$1$ share through
port~$1$. 

Note that also here, still only one \emph{single} test packet 
will be 
sent to collect the 
capacity information. This information will be stored 
in the packet-variable~$c$ as well. However, when the test 
packet arrives at a switch where a flow can be split,
copies of the packet are sent (after updating the~$c$ according to the 
bandwidth of each path) to all possible paths, 
to record the capacity along all other paths. This exploits
the fact that~$\system$ (NetKAT) treats the~$\&$ operator 
as \emph{conjunction} in the sense that both operations are performed, 
rather than \emph{disjunction}, where one of the two operations 
would be chosen non-deterministically (according to the
usual Kleene interpretation). 
Again, we emphasize that 
we will refer to~$c$ stored in one \emph{single} test packet,
and not the
actual real data flow.
 Now the topology will update $c$ as in the unsplittable case. 
However, the policy needs to not only decide which ports the packets 
go to, but also update~$c$ according to the split policy.  
E.g.,, at~$dc_2$, the data flow from~$dc_1$ to~$dc_5$ 
at rate~$4$ is sent out through port~$1$ at rate~$3$, 
and the port~$3$ at~$1$. And if the rate is smaller
 than or equal to~$3$, e.g.,~$2$, then the whole flow of 
rate~$2$ will be sent out through port~$1$.
The following~$\system$ formula specifies this behavior:
\\

\centerline{
$\begin{array}{lcl}
src=dc_1;dst=dc_5;sw=dc_2;pt=4;c\leq 5\\
\phantom{sw=}(pt\leftarrow 1;c\leftarrow \min\{3,c\}\\
\phantom{sw=}\&\ pt\leftarrow 3;c\leftarrow \max\{0,c-3\})
\end{array}$
}

The test~$c\leq 5$ ensures
that the flow does not exceed the 
capacity of both paths. Notice that even when the size 
of the flow is small enough for one path, a copy of the test packet 
with~$c=0$ will still be sent to the other. 
This ensures that 
sufficient information  is available 
at the switch where flows merge.
That is, 
the switch collects the weights the packets carry 
($c$ in our example). 
The switch will only push packets to the right out-ports 
after all expected packets have arrived. 
This will happen before the switch sends the packet to the right out-ports.
For example, at~$dc_4$, the flow from~$dc_1$ to~$dc_5$ might 
arrive in from ports~$1$ and~$2$ and will be sent out through port~$3$. 
In order to record the capacity of both links, switch-variables~$C$ and~$X$ 
are introduced, for each possible merge. For example, 
the following table provides the merging rules for 
the switch at~$dc_4$, 
where~$X$ is the counter for the merge, and~$C$ stores the current 
capacity of the arriving test packets. Initially,~$X$ is set to the 
number of in-ports for the merge, and~$C$ is set to 0.  

\centerline{$\begin{array}{ccccccccccc}
\text{src}&\phantom{ee}&\text{dst}&
\phantom{fe}&\text{in}&\phantom{ee}&\text{out}&\phantom{ee}&
{C}&\phantom{ee}&X\\
dc_1&&dc_5&&1,2&&3&&0&&2\\
dc_5&&dc_2&&3,4&&1,2&&0&&2
\end{array}$
}

The first line of the rules in the table can be specified in~$\system$ as follows:




\centerline{
$\begin{array}{c}
sw=dc_4;src=dc_1;dst=dc_5;(pt=1\ \&\ pt=2);\\C\leftarrow C+c;X\leftarrow X-1;\\
(X\not=0;drop\ \&~~ X=0;c\leftarrow C; pt \leftarrow 3)
\end{array}$~
}


When a packet from~$dc_1$ to~$dc_5$ arrives 
at port~$1$ or~$2$ of~$dc_4$, first the switch collects
the value of~$c$ and adds it to the switch-variable~$C$, 
then decrements~$X$ to record that one packet arrived. 
Afterwards, we test whether all expected packets arrived ($X=0$).
If not, the current one is dropped; 
if yes, we send the current packet out to port~$3$. 
The reason that we can drop 
all packets except for the last, is that all those packets carry exactly the 
same values. 
Therefore, we eventually only need to include 
 the merged capacity ($C$) in the last 
packet, and propagate it.

Combining the split and merge cases, the policy of 
the switch can be defined. For example, 
the second line of the merging rule table can be specified 
as follows, by first merging from port~$3$ and~$4$, 
and then splitting to port~$1$ and~$2$:

\centerline{$\begin{array}{l}
sw=dc_4;src=dc_5;dst=dc_2;(pt=3\ \&\ pt=4);\\
\phantom{sw=}C\leftarrow C+c;X\leftarrow X-1;\\
\phantom{sw=}(X\not=0;drop\ \&\ X=0;c\leftarrow C; c\leq 8\\
\phantom{sw=X\not=0;drop\ \&\ }(pt\leftarrow 1;c\leftarrow \min\{6,c\}\\
\phantom{sw=X\not=0;drop\ \&\ }\&\ pt\leftarrow 2;c\leftarrow \max\{0,c-6\}))
\end{array}$
}

Then the splittable capacited reachability question 
can be answered by checking whether the following 
expression evaluates to~$drop$:
\\\centerline{$\begin{array}{c}
scr\leftarrow A;dst\leftarrow B;c\leftarrow r;sw\leftarrow A;pt(pt)^\ast;\\
sw=B;X=0;c\geq r
\end{array}$
}

If the above formula does not equal~$drop$, 
then~$A$ can communicate at rate at least~$r$ to~$B$.

\end{example}

 \subsection{Service Chaining}

The virtualization and programmability trend is not limited
to the network, but is currently also discussed intensively
for network functions in the context of the Network
Function Virtualization (NFV) paradigm. 
SDN and NFV nicely complement
each other, enabling innovative new network services such
as \emph{service chains}~\cite{servicechains}: network functions which are
traversed in a particular order (e.g., first firewall, then cache,
then wide-area network optimizer).
Our language allows to reason about questions
such as \emph{Are sequences of network functions traversed
in a particular order, without violating node and link capacities?}
$\system$ can easily be used to describe weighted aspects also
in the context of service chains. 
In particular, network functions may both \emph{increase}
(e.g., due to addition of an encapsulation header, or
a watermark) or \emph{decrease} (e.g., a WAN optimizer,
or a cache) the traffic rate, both \emph{additively} (e.g.,
adding a header) or \emph{multiplicatively} (e.g., WAN optimizer). 

\begin{example} 
Let us go back to Figure~\ref{fig:sc},
and consider a service chain of the form~$(s,F_1,F_2,t)$: traffic from
$s$ to~$t$ should first traverse a function~$F_1$ and then a function~$F_2$,
before reaching~$t$. For example,~$F_1$ may be a firewall or proxy
and~$F_2$ is a WAN optimizer. The virtualized functions~$F_1$ and
$F_2$ may be allocated redundantly and may change the traffic volume.
Using~$\system$, we can ask questions such as: \emph{What is the maximal
rate at which~$s$ can transmit traffic into the service chain?}
or  \emph{Can we realize a service chain of cost (e.g., latency) at most~$x$?}.
Let us consider the following example:
The question \emph{``Can~$s$ reach~$t$ at cost/latency at most~$\ell$ and/or at rate/bandwidth at least~$r$, via the service chain functions~$F_1$ and~$F_2$?''}, can be formulated by combining the reachability problems above and the 
waypointing technique in~\cite{netkat}. For example,
in case of cost reachability, 
we can ask if the following~$\system$ formula equals~$drop$. 
\\\centerline{$\begin{array}{c}src\leftarrow s;dst\leftarrow t;co\leftarrow 0;sw\leftarrow s;
pt(pt)^\ast;\\
sw=F_1;p_{F_1};tpt(pt)^\ast;sw=F_2;p_{F_2};\\
tpt(pt)^\ast;sw=t;co\leq \ell;ca\geq r
\end{array}$}

Note that in this example, we considered an unsplittable 
scenario. For the splittable scenario, 
we can extend the splittable capacitated reachability 
use case above analogously.  
\end{example}

\subsection{Fairness}

Related to quality-of-service is the question of fairness.
For example, a natural question to ask is: 
\emph{``Does the current flow allocation satisfy 
network neutrality requirements?''}~\cite{net-neutral}, or more specifically,
\emph{``Is the network max-min fair?''}~\cite{maxmin}

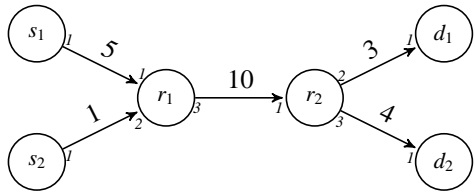
\begin{figure}[!htb]
\vspace{-.8em}
\begin{center}
\begin{tikzpicture}[->,>=stealth',shorten >=1pt,auto,
                    semithick,scale=0.85, every node/.style={scale=0.85}]
  \tikzstyle{every state}=[fill=white,draw,text=black]
 \tikzstyle{vecArrow} = [thick, decoration={markings,mark=at position
   1 with {\arrow[semithick]{open triangle 60}}},
   double distance=1.4pt, shorten >= 5.5pt,
   preaction = {decorate},
   postaction = {draw,line width=1.4pt, white,shorten >= 4.5pt}]
\tikzstyle{innerWhite} = [semithick, white,line width=1.4pt, shorten >= 4.5pt]
    \node[state] (s1) at (0,0) {$s_1$};    
    \node[state] (s2) at (0,-2) {$s_2$};
    \node[state] (B1) at (2,-1) {$r_1$};
    \node[state] (B2) at (4.3,-1) {$r_2$};
    \node[state] (E1) at (6.3,0) {$d_1$};
    \node[state] (E2) at (6.3,-2) {$d_2$};
    \path[every node/.style={sloped,anchor=south,auto=false,scale=1}]
	 (s1) edge node[midway,above] {$5$} (B1)
	 (s2) edge node[midway,above] {$1$} (B1)
	 (B1) edge node[midway,above] {$10$} (B2)            
	 (B2) edge node[midway,above] {$3$} (E1)     
	      edge node[midway,above]  {$4$} (E2);
	\node (pt1) at (0.5,-0.1) {\scriptsize \textit{1}};
	\node (pt1') at (1.63,-0.63) {\scriptsize \textit{1}};
	\node (pt2) at (0.5,-1.9) {\scriptsize \textit{1}};	
	\node (pt2') at (1.58,-1.4) {\scriptsize \textit{2}};
	\node (pt3) at (2.48,-1.15) {\scriptsize \textit{3}};
	\node (pt3') at (3.75,-1.15) {\scriptsize \textit{1}};	
	\node (pt4) at (4.72,-0.65) {\scriptsize \textit{2}};
	\node (pt5) at (4.69,-1.38) {\scriptsize \textit{3}};	
	\node (pt4') at (5.77,-0.09) {\scriptsize \textit{1}};
	\node (pt5') at (5.78,-1.93) {\scriptsize \textit{1}};	
\end{tikzpicture}
\end{center}
\vspace{-1em}
\caption{Max-min Fairness}\label{fig-fair}
\vspace{-1em}
\end{figure}

For example, consider the network in 
Figure~\ref{fig-fair}. The numbers on the links 
specify the bandwidth capacity. Suppose that
there are three flows:~$s_1\rightarrow d_1,s_1\rightarrow d_2,s_2\rightarrow d_2$ 
embedded in this network. 
Suppose the rates of these three flows 
are~$2,3,1$, respectively. 
In a max-min fair allocation, we aim to maximize
the minimal flow allocated to any
of these three flows, subject to 
capacity constraints. For example, the minimum flow 
$s_2\rightarrow d_2$ receives a fair share here:
the flow is naturally limited by the first link of
capacity 1.
However, the next smallest rate, $s_1\rightarrow d_1$,
may be increased to $2.5$, by reducing the flow
$s_1\rightarrow d_2$ accordingly.

\begin{example}
Consider the network and the flows mentioned earlier in Figure~\ref{fig-fair}. 
In order to specify whether the given rates for the flows are max-min fair, 
we send three test packets along the three flows respectively. 
The topology of this network can be specified in~$\system$ as before, 
which only takes care of forwarding packets. E.g., the link between~$s_1$ 
and~$r_1$ is specified as follows, where~$c$ specifies the capacity:
\\\centerline{$sw=s_1;pt=1;sw\leftarrow r_1;pt\leftarrow 1$}

The switches not only put incoming packets to the right 
outgoing ports, but also assign the bandwidth 
of the out-going links to different flows. 
This will be recorded in the test packets as a packet-variable, 
denoted by~$a$. Moreover, three packet-variables~$x_1,x_2,x_3$ 
are used to represent the rates of the flows that are 
supposed to be checked for max-min fairness.

E.g., the policy of~$r_1$ can be specified as follows.
\\\centerline{$\begin{array}{l}
\ \ sw=r_1;(pt=1\ \&\ pt=2);pt\leftarrow 3;\\
\phantom{sw=r_1\&}(scr=s_1;dst=d_1;a\leftarrow\min\{x_1,a,c-x_2-x_3\}\\
\phantom{sw=r_1}\&\ scr=s_1;dst=d_2;a\leftarrow\min\{x_2,a,c-x_1-x_3\}\\
\phantom{sw=r_1}\&\ scr=s_2;dst=d_2;a\leftarrow\min\{x_3,a,c-x_1-x_2\}
\end{array}
$}

The first line specifies that the packets arriving at 
port~$1$ or~$2$ will be sent out through 
port~$3$. The following lines update~$a$ 
of the test packets, according the rates of the 
flows that share the link. 
In order to test max-min fairness, 
we first order the given rates increasingly. 
For example, to check whether the following flow rates 
are max-min fair,
\\\centerline{$x_1=2,x_2=3,x_3=1$}   
we represent them as~$x_3=1,x_1=2,x_2=3$. 
Then, we check them one-by-one, 
by verifying whether the following~$\system$ expressions
are equal to~$drop$ or not.
\\\centerline{$\begin{array}{lcl}
f_3&=&sw\leftarrow s_2;scr\leftarrow s_2;dst\leftarrow d_2;a\leftarrow 10;\\
&&x_3\leftarrow 1;x_1\leftarrow 0;x_2\leftarrow 0;tp(tp)^\ast;sw=d_2;x_3=a\\
f_1&=&sw\leftarrow s_1;scr\leftarrow s_1;dst\leftarrow d_1;a\leftarrow 10;\\
&&x_1\leftarrow 2;x_3\leftarrow 1;x_2\leftarrow 0; tp(tp)^\ast;sw=d_1;x_1=a\\
f_2&=&sw\leftarrow s_1;scr\leftarrow s_1;dst\leftarrow d_2;a\leftarrow 10; \\
&&x_2\leftarrow 3;x_1\leftarrow 2;x_3\leftarrow 1; tp(tp)^\ast;sw=d_1;x_2=a\\
\end{array}
$}

While checking whether~$x_3$ is max-min fair, 
one does not need to consider~$x_1,x_2$ (can be modified if not fair), 
because~$x_3$ is the minimum; therefore,~$x_1,x_2$ are set to~$0$.  
While checking~$x_1$, it is already known 
that~$x_3$ is fair. Therefore the value of~$x_3$ cannot be 
changed even if~$x_1$ is not fair; 
this is why~$x_3$ is set to~$1$, and~$x_2$ is~$0$ in~$f_1$. 
Similarly for~$x_2$.

If all of the above formulas are not equal to drop, 
then the given rates are max-min fair. E.g., 
the above rates~$x_1=2,x_2=3,x_1=1$ are not max-min fair 
because~$f_1$ equals~$drop$. This matches our discussion above.

Moreover, the results of the above tests 
can also help develop more fair rates. 
E.g.,~$f_1$ is the first list entry to equal~$drop$,
this means that~$x_1$ is not fairly allocated. 
However,~$x_3$ must be fair as $f_3$ does not equal~$drop$. 
Therefore, to ensure max-min fairness in this network, 
one should keep the rate of the flow~$s_2\rightarrow d_2$, 
and adjust the rate of~$s_1\rightarrow d_1$ and that of~$s_1\rightarrow d_2$.
\end{example}

\subsection{Quality-of-Service}

We have already shown how to specify and reason about 
a number of 
relevant use cases for weighted networks. 
An additional important motivation for weighted
models concerns the ability to express and verify
\emph{quality-of-service} guarantees.
For example, in a computer network (providing
limited resources), it can be useful
to prioritize a certain flow (e.g., a VoIP call) over
another (e.g., a Dropbox synchronization).

\begin{example}\label{exm-qos}
Consider the switch in Figure \ref{fig-qos}. 

\begin{minipage}{0.25\textwidth}
\begin{figure}[H]
\vspace{-0.8em}
\begin{center}
\begin{tikzpicture}[->,>=stealth',shorten >=1pt,auto,
                    semithick,scale=0.75, every node/.style={scale=0.75}]
  \tikzstyle{every state}=[fill=white,draw,text=black]
 \tikzstyle{vecArrow} = [thick, decoration={markings,mark=at position
   1 with {\arrow[semithick]{open triangle 60}}},
   double distance=1.4pt, shorten >= 5.5pt,
   preaction = {decorate},
   postaction = {draw,line width=1.4pt, white,shorten >= 4.5pt}]
\tikzstyle{innerWhite} = [semithick, white,line width=1.4pt, shorten >= 4.5pt]
    \node (s1) at (0,0) {};    
    \node (s2) at (0,-2) {};
    \node[state] (B1) at (2,-1) {$r$};
    \node (B2) at (4.3,-1) {};
    
    \path[every node/.style={sloped,anchor=south,auto=false,scale=1}]
	 (s1) edge node[midway,above] {} (B1)
	 (s2) edge node[midway,above] {} (B1)
	 (B1) edge node[midway,above] {} (B2)  ;
	 
	\node (pt1') at (1.6,-0.65) {\scriptsize \textit{1}};	
	\node (pt2') at (1.56,-1.38) {\scriptsize \textit{2}};
	\node (pt3) at (2.5,-1.15) {\scriptsize \textit{3}};
	
	\node (x) at (2,-1.8) {$C_h,C_l$};
\end{tikzpicture}
\end{center}
\vspace{-1.2em}
\caption{QoS}\label{fig-qos}
\vspace{0.5em}
\end{figure}
\end{minipage}\hfill
\begin{minipage}{0.7\textwidth}
It has two incoming ports $1,2$ and one outgoing port $3$. 
Suppose there are two types of traffic flows going through this switch: 
skype calls and web surfing traffic. This switch should respect that the skype 
calls have higher priority, but at the same time not completely disallow 
the web surfing. 
Suppose we want to give skype calls 80\% of the bandwidth and to web 
surfing only 20\%.  This policy can be easily specified with
\end{minipage}

\noindent the following \system 
expression, where $x$ is a packet-variable specifying the priority of the packet 
(e.g., skype has high priority $high$ and http low $low$) and $C_h,C_l$ are two switch-variables for counting the number of packets with the two priorities respectively. 
\\\centerline{$\begin{array}{l}
sw=r;(pt=1\ \&\ pt=2);\\
\phantom{sw=r\ }(x=high; C_h<8;pt\leftarrow 3;C_h\leftarrow C_h+1;\\
\phantom{sw=r}\&\ x=low; C_l<2;pt\leftarrow 3;C_l\leftarrow C_l+1);\\
\phantom{sw=r}C_h=8;C_l=2;C_h\leftarrow 0;C_l\leftarrow0
\end{array}$}

The second line  deals with high priority packets: if the amount of the 
packets with this priority is less than $8$, then it will be sent out through 
port $3$. Similar for the low priority packets. The last line resets $C_h,C_l$ 
to $0$ when both reach the upper limit, 
triggering a new round of counting.
\end{example}


\subsection{Further Extensions}\label{sec:extensions}

While, using switch-variables
(e.g., as counters), 
$\system$  
supports a basic form of prioritization,
allowing to provide one flow 
with a larger share of the
bandwidth than another, this solution is still naive. 
For example, when a skype packet arrives at switch $r$ and $C_h$ 
is already $8$, then this packet will be dropped, which might lower the 
quality of the skype call. In a even worse situation, in the absence
of web traffic packets, the switch will still wait and drop all the 
incoming skype packets. 
To overcome these problems
and improved notion of quality-of-service, we could introduce a notion of
 \emph{queue}.
Indeed, queues, e.g., annotated with different round robin
weights, are an essential component in any computer network today, and are also the predominant mechanism to implement 
service differentiation.
However, while OpenFlow actions can readily be used to \emph{enqueue} 
a packet in a certain queue, it is the responsibility of the management 
plane (and not the SDN control plane) to actually allocate these 
queues and scheduling policies. While we currently witness first attempts to 
combine control and management planes~\cite{mergeplane},
 today, there does not exist any 
standard. 
Nevertheless, we in the following start exploring how~$\system$ 
could be extended with a notion of queues. 

Concretely, we can extend~$\system$ by a set of queues, 
henceforth denoted by~$\mathbf{Q}$, which are used for 
buffering packets which currently cannot be forwarded 
due to limited resources.
We will assume that all queues are FIFO with normal queue related functions,
 e.g., enqueue ($EQ(\ )$), dequeue ($DQ(\ )$), head of queue ($HEAD(\ )$), etc.


For specifying the queue operations, we extend~$\system$ 
to allow \emph{enqueue} and \emph{dequeue}, where~$\qu\in\mathbf{Q}$:
\begin{itemize}
\item \textbf{Enqueue}~$\textsf{EQ } \qu$: Put the current packet into the 
queue~$\qu$. The packet remains in the queue until being processed by the switch.
\item \textbf{Dequeue}~$\textsf{DQ } \qu$: Dequeue the first packet from the 
queue~$\qu$ and delete it from the queue.
\end{itemize} 

The semantics are defined in Table~\ref{sem-2}, for~$\qu\in\mathbf{Q}$.

\begin{table}[]
$$\begin{array}{rcl}
\llbracket\textsf{EQ }\qu\rrbracket(\rho,\ pk::h)&=&
\left\{\begin{array}{lcl}
\llbracket 1^\ast\rrbracket(\rho,\ pk::h)&&\text{if }\qu\not= FULL,\\
&&\text{then }EQ(\qu)\\
\emptyset&&\text{otherwise}
\end{array}\right.
\hfill (5)\\\\
\llbracket\textsf{DQ }\qu\rrbracket(\rho,\ pk::h)&=&\left\{\begin{array}{lcl}
\{\rho,\ pk::h\}&\text{if }HEAD(\qu)=pk::h\\
&\text{then }DQ(\qu)\\
\llbracket 1^\ast\rrbracket(\rho,\ pk::h)&\text{otherwise}
\end{array}\right.\hfill (6)
\end{array}$$
\caption{Semantics for Queuing}\label{sem-2}
\vspace{-2em}
\end{table}

\emph{Remarks:}  Equations (5) and (6) deal with the queues at the switches, by taking care of the detailed queue processing. Notice that each switch can only manage its own 
queues. \hfill~$\blacksquare$

\begin{example}

Consider the same switch in Example \ref{exm-qos}. 

\begin{minipage}{0.25\textwidth}
\begin{figure}[H]
\vspace{-0.8em}
\begin{center}
\begin{tikzpicture}[->,>=stealth',shorten >=1pt,auto,
                    semithick,scale=0.75, every node/.style={scale=0.75}]
  \tikzstyle{every state}=[fill=white,draw,text=black]
 \tikzstyle{vecArrow} = [thick, decoration={markings,mark=at position
   1 with {\arrow[semithick]{open triangle 60}}},
   double distance=1.4pt, shorten >= 5.5pt,
   preaction = {decorate},
   postaction = {draw,line width=1.4pt, white,shorten >= 4.5pt}]
\tikzstyle{innerWhite} = [semithick, white,line width=1.4pt, shorten >= 4.5pt]
    \node (s1) at (0,0) {};    
    \node (s2) at (0,-2) {};
    \node[state] (B1) at (2,-1) {$r$};
    \node (B2) at (4.3,-1) {};
    
    \path[every node/.style={sloped,anchor=south,auto=false,scale=1}]
	 (s1) edge node[midway,above] {} (B1)
	 (s2) edge node[midway,above] {} (B1)
	 (B1) edge node[midway,above] {} (B2)  ;
	 
	\node (pt1') at (1.6,-0.65) {\scriptsize \textit{1}};	
	\node (pt2') at (1.56,-1.38) {\scriptsize \textit{2}};
	\node (pt3) at (2.5,-1.15) {\scriptsize \textit{3}};
	
	\node (y) at (2,-0.3) {$\qu_h,\qu_l$};
	\node (x) at (2,-1.8) {$C_h,C_l$};
\end{tikzpicture}
\end{center}
\vspace{-1.2em}
\caption{QoS}\label{fig-qqos}
\vspace{0.5em}
\end{figure}
\end{minipage}\hfill
\begin{minipage}{0.7\textwidth}
However, there are two queues at the switch for high priority packets (e.g., skype packets) and low priority (e.g., http packets) packets respectively. 

Different from Example \ref{exm-qos}, when a packet arrives at the switch $r$, it will be put into the right queue first. 
This can be specified using the following, where $x$ is the packet-variable representing the priority.
\end{minipage}

\centerline{$\begin{array}{l}
\ sw=r;x=low;\textsf{EQ }\qu_l \ \&\ sw=r;x=high;{\textsf{EQ }\qu_h}
\end{array}
$}

Moreover, the switch also makes sure that the high priority 
queue is processed 80\% of the time and the low priority queue 20\% of the time. The following expression shows the case of high priority packets.
\\\centerline{$\begin{array}{l}
sw=r;x=high;\\
\phantom{sw-}(x_h<8;{\textsf{DQ }\qu_h};pt\leftarrow 3;x_h\leftarrow x_h+1\\
\phantom{sw-}\&\ x_h=8;Q_l=\emptyset;{\textsf{DQ }\qu_h};pt\leftarrow 3\\
\phantom{sw-}\&\ x_h=8;Q_l\not=\emptyset;skip)\\
\phantom{sw-}x_h=8;x_l=2;x_h\leftarrow 0;x_l\leftarrow 0
\end{array}
$}

The second line specifies that when $x_h$ is less than $8$, 
we take the first packet of the high priority queue and send it through port $3$. 
This is similar to non-queue case.
The third line specifies the situation when $x_h$ already reached 
its upper limit, meaning that the high priority packets already used up the 
bandwidth assigned for them. However, the low priority queue is empty, 
i.e., there is no low priority packet that needs to use the link. Therefore, in this case, 
the high priority packets can use the low priority packets' share. 
The fourth line specifies the case that the high priority packets already used up
their share and need to wait for the low packets to go first. 
The last line tests whether the counters both reach their upper limit and if yes,  
reset both of them.
\end{example} 
 
Related to the quality-of-service discussion above
is also the question of how to model
entire flows competing for a \emph{set} of shared resources
(e.g., along paths).
While so far, all our use cases have been described in terms
of packet and switch variables only, these concepts are insufficient
to model contention across
multiple resources.

In principle, it is simple to extend~$\system$ with a notion
of global variables which allows to account for such more global
aspects. 
In practice however, supporting global variables can
be inefficient: such variables cannot be maintained
by the switch, but require interactions with the controller.
The latter introduces network loads and latencies,
which can be undesirable, especially in wide-area network
where the controller can be located remotely.


\section{(Un)Decidability}\label{sec:deci}

In this section we shed light on the
fundamental decidability of weighted SDN programming
languages like WNetKAT. 
Given today's trend toward more quantitative
networking, we believe that this is an
important yet hardly explored dimension.
In particular, we will establish an equivalence
between WNetKAT and weighted automata.

In the following, we will restrict 
ourselves to settings where quantitative variables
of the same type
behave similarly in the entire network:
For example, the cost variables (e.g., quantifying latencies) in the 
network are always added up along a given path, while
capacity variables require minimum operations 
along different paths. This is a reasonable 
for real-world networks.

The definition of the weighted automata used here is slightly different
from those usually studied, e.g.,~\cite{DBLP:conf/icalp/DrosteG05,droste2009handbook}. 
However, it is easy to see that they are equivalent. 


We first introduce some preliminaries.
A \emph{semiring} is a structure $(K, \oplus, \otimes, 0, 1)$, 
where $(K, \oplus, 0)$ is a commutative
monoid, $(K, \otimes, 1)$ is a monoid, multiplication distributes 
over addition $k\otimes (k'\oplus k'')=k\otimes k'\oplus k\otimes k''$, 
and $0\otimes k =
k\otimes 0 = 0$ for each $k \in K$. 
For example, $(\naturals\cup\{\infty\}, \min, +,\infty, 0)$ 
and $(\naturals\cup\{\infty\}, \max, +,\infty, 0)$ are semirings, 
named the \emph{tropical semiring}. $(\naturals\cup\{\infty\}, \max, \min,0,\infty)$ 
is also a semiring.
A \emph{bimonoid} is a structure $(K, \oplus, \otimes, 0, 1)$, where $(K, \oplus, 0)$ and $(K, \otimes, 1)$ are monoids. $K$ is called a \emph{strong bimonoid} if $\oplus$ is commutative and $0\otimes k =k\otimes 0 = 0$ for each $k \in K$.
For example, $(\naturals\cup\{\infty\},+, \min,0,\infty)$ is a (strong) bimonoid, named the \emph{tropical bimonoid}.

Now fix a semiring/bimonoid $K$ and an alphabet $\Sigma$.
A \emph{weighted finite automaton} (WFA) over $K$ and $\Sigma$ is a 
quadruple $A = (S, s,F, \mu)$ where
$S$ is a finite set of states, $s$ is the starting state, $F$ is set of the final states,  $\mu : \Sigma\to K^{S\times S}$ is the transition weight function and $\lambda$ is the weight of entering the automaton. For $\mu(a)(s,s')=k$, we write $s\tr{a}{k}s'$.

Let $\sf At$ be the set of complete non-quantitative tests 
and $P$ be the set of  complete non-quantitative assignments.
Let $\Omega$ be the set of complete quantitative tests and 
$\Delta$ be the set of complete quantitative assignments.


A weighted NetKAT automata is a finite state weighted automaton 
$A = (S, s,F,\lambda, \mu)$ over a structure $K$ and alphabet $\Sigma$. 
The inputs to the automaton are so called reduced strings introduced in \cite{netkat,coalg}, which belong to the set $U=\sf{At}\cdot\Omega\cdot P\cdot\Delta\cdot(\sf{dup}\cdot P\cdot\Delta)^\ast$, i.e., the strings belonging to $U$ are of the form: 
$$\alpha\omega p_0\delta_0\ \sf{dup}\ p_1\delta_1\ \sf{dup}\ \cdots\ \sf{dup}\ p_n\delta_n$$
for some $n\geq 0$. Intuitively, $\mu$ attempts to consume $\alpha\omega p_0\delta_0\ \sf{dup}$ from the front of the input string and move to a new state with a weight and the new state has the residual input string $\alpha_0\omega_{0}\ p_1\delta_1\ \sf{dup}\ \cdots\ \sf{dup}\ p_n\delta_n$. 

\newpage

The following construction shows the equivalence between 
WNetKAT and weighted automata.

{\bf From WFA to WNetKAT}

Let $A = (S, s, F,\lambda,\mu)$ be a weighted NetKAT automata over $K$ and $\Sigma$. An accepting path in $A$ $s\tr{r_1}{\alpha_1\beta_1}s_1\tr{r_2}{\alpha_2\beta_2}s_2\cdots\tr{r_n}{\alpha_n\beta_n}s_n$ can be write as the following WNetKAT expression:

$\alpha_1\omega_1 p_1\delta_1\ {\sf dup}\ p_2\delta_2\ {\sf dup}\ \cdots\ {\sf dup}\ p_n\delta_n$, where 
\\1. $\omega_1=\lambda$, $\delta_1=\omega_1\oplus r$ and $\delta_{i}=\delta_{i-1}\oplus r_{i}$ for $i=2,...,n$;
\\2. $p_i=p_{\beta_i}$ for $i=1,...,n$.

{\bf From WNetKAT to WFA}

Let $e$ be a weighted automata expression, then following \cite{netkat,coalg}, we can define a set of reduced strings $R$ which are semantically equivalent to $e$. We define a weighted NetKAT automata $A = (S, s,F,\lambda, \mu)$ over a structure $K$ and alphabet $\Sigma$, where 
\\
$s=R$ and $\Sigma={\sf At}\times{\sf At}$. 
\\
$\mu: \Sigma\to K^{S\times S}$ is defined as: 
$\mu(\alpha,\beta)(u_1,u_2)=r$ iff $u_2=\{\beta\omega' x\mid\alpha\omega p\delta\ {\sf dup}\ x\in u_1\}\text{, where }\beta=\alpha_p, \omega'=\delta_\omega\text{ and }\omega\otimes r=\omega'.$ For short write $u_1\tr{r}{\alpha\beta}u_2$.
\\
$S=\{s\}\cup\{u\subseteq 2^U\mid\exists\ \mu\text{-path }s\to\cdots\to u\}$
\\$F=\{u\mid\alpha\omega p\delta\in u\in S\}$
\\$\lambda=\{\omega\mid\alpha\omega x\in s\}$

We have the following theorem.
\begin{theorem}
(1) For every finite weighted WNetKAT automaton $A$, there exists a WNetKAT expression $e$ such that the set of reduced strings accepted
by $A$ is the set of reduced strings of $e$.
(2)~For every WNetKAT expression $e$, there is a weighted WNetKAT automaton $A$ accepting the set of the reduced strings of $e$.
\end{theorem}

Let us just give some examples:
\begin{enumerate}
\item For the cost reachability use case, there exists a weighted WNetKAT automaton over the tropical semiring $(\naturals\cup\{\infty\}, +, \min,\infty,0)$ that accepts the set of  reduced strings of the WNetKAT expression in Section~\ref{sec:app-cost}.

\item For the capacitated reachability: (i)
There exists a weighted WNetKAT automaton over the semiring $(\naturals\cup\{\infty\}, \max, \min,0,\infty)$ that accepts the set of the reduced strings of the WNetKAT expression for the splitable case in Section \ref{sec:app-cap}.
(ii) There exists a weighted WNetKAT automaton over the tropical bimonoid $(\naturals\cup\{\infty\}, \min,+,0,\infty)$ that accepts the set of the reduced strings of the WNetKAT expression for the unsplitable case in Section~\ref{sec:app-cap}.
\end{enumerate}

From this relationship, we have the following theorem about the (un)decidability of WNetKAT expression equivalence.
\begin{theorem}
Deciding equivalence of two WNetKAT expressions is equal to deciding the equivalence of the two corresponding weighted WNetKAT automata.
\end{theorem}

For all the semiring and bimonoid we encountered in this paper, the WFA equivalence is undecidable. Therefore, the equivalence is also undeciable. 

This negative result highlights the inherent
challenges involved in complex network languages
which are powerful enough to deal with weighted aspects.

However, we also observe that in many practical scenarios,
the above undecidability result is too general and does not apply.
For example, most of the use cases presented in
in Section \ref{sec:applications} can actually be reduced to
test \emph{emptiness}: we often want to test 
whether a given WNetKAT expression $e$ equals $0$, i.e., 
whether the corresponding weighted NetKAT automaton is empty. Indeed, there seems to exist an intriguing relationship between
emptiness and reachability. 

\begin{theorem}
Deciding whether a WNetKAT expression is equal to $0$ is equal to deciding the emptiness of the corresponding weighted automaton.
\end{theorem}

Interestingly, as shown in~\cite{DBLP:conf/ncma/DrosteG13,DBLP:journals/fuin/DrosteH15,DBLP:conf/dlt/Kirsten09,DBLP:journals/ipl/KirstenQ11}, 
the emptiness problem is decidable for several semirings/bimonoids, 
e.g., the tropical semiring and the tropical bimonoid used in this paper. This leads to the decidability of the WNetKAT equivalence over these structures.

Another interesting domain with many decidability
results are unambiguous regular grammars
and unambiguous finite automata~\cite{unamb1}.
Accordingly, in our future work,
we aim to extend these concepts to the weighted world
and explore the unambiguous subsets
of WNetKAT which might enable decidability for equivalence. 
 
\section{Compilation and Practical Considerations}\label{sec:compilation}

We conclude with some remarks on compilation
and compatibility to OpenFlow.  
In general, OpenFlow today does not accommodate 
\emph{stateful} packet operations,
and thus, per-connection or per-packet logic
require involvement of the controller. 
Moreover, OpenFlow switches do not per se support arithmetic 
computations, such as addition of packet field values. 

However, we currently witness a strong trend toward
computationally more advanced and stateful
packet-processing functionality.
For example, the advent of 
programming protocol-independent packet processors
like P4~\cite{p4}, 
programming platform-independent stateful OpenFlow 
applications inside the switch
like OpenState~\cite{openstate},
but also systems like SNAP~\cite{snap}, POF~\cite{pof},
or Domino~\cite{domino} introduce features
which render these platforms potentially interesting 
compilation targets for $\system$. 
To give another example, Open vSwitch 
allows running on-hypervisor “local
controllers” in addition to a remote, primary controller,
to introduce a more stateful packet processing.

Nevertheless, we observe that several features 
of \emph{today's} OpenFlow versions can be exploited
for the compilation of $\system$ expressions.
For example, in order to
implement arithmetic operations (see e.g., Equations (3) and (4)), 
we can simply use lookup tables realized as
OpenFlow rules, see the technique in~\cite{brain}. Thus, cost reachability 
queries can be compiled to flow tables similarly as in NetKAT.

Interestingly, however, also
a simple form of switch state can readily be implemented
in OpenFlow today. Indeed, OpenFlow switches
provide stateful features such as group tables, packet counters,
bandwidth meters, etc. 
For example, a simple yet inefficient solution to compile
$\system$ switch variables is to use either the 
meta-rule approach taken by Schiff 
et al.~\cite{ccr16sync} (introducing
additional flow rules), or to leverage round
robin groups~\cite{brain}. 
Finally, we note that while 
OpenFlow actions can be used to forward
packets to specific queues, the scheduling regime for the queues 
is defined via the
management plane~\cite{mergeplane}.

\section{Conclusion}\label{sec:conclusion}

In our future research, we aim to chart a more comprehensive
landscape of the decidability and decision complexity of~$\system$. 
In the longer term, we also aim to extend~$\system$
to support probabilistic aspects of networking.

\section*{Acknowledgements}

We would like to thank Alexandra Silva, Nate Foster, 
Dexter Kozen, Manfred Droste and Fredrik Dahlqvist for 
many inputs and discussions on
WNetKAT.


\bibliographystyle{abbrv}

\end{document}